\documentclass[useAMS,usenatbib,usegraphicx]{mn2e}

\usepackage{amsmath}

\title[HOD of massive galaxies since $z = 1$]
{Halo Occupation Distribution of Massive Galaxies since $z = 1$}
\author[Y. Matsuoka et al.]{Y. Matsuoka$^{1,2}$\thanks{Research Fellow of the Japan Society for the Promotion 
of Science}\thanks{E-mail: matsuoka@a.phys.nagoya-u.ac.jp}, S. Masaki$^{1}$, K. Kawara$^{2}$, and N. Sugiyama$^{1,3}$\\
$^{1}$Graduate School of Science, Nagoya University, Furo-cho, Chikusa-ku, Nagoya 464-8602, Japan\\
$^{2}$Institute of Astronomy, The University of Tokyo, Osawa 2-21-1, Mitaka, Tokyo 181-0015, Japan\\
$^{3}$Institute for the Physics and Mathematics of the Universe, The University of Tokyo, Kashiwanoha, Kashiwa, Chiba 277-8568, Japan}

\begin{document}

\date{Accepted ---. Received ---; in original form ---}

\pagerange{\pageref{firstpage}--\pageref{lastpage}} \pubyear{2010}

\maketitle

\label{firstpage}

\begin{abstract}
We present a clustering analysis of $\sim$60,000 massive (stellar mass $M_{\star} > 10^{11} M_{\odot}$) galaxies out to $z = 1$
drawn from 55.2 deg$^2$ of the UKIRT Infrared Deep Sky Survey (UKIDSS) and the Sloan Digital Sky Survey (SDSS) II Supernova Survey.
Strong clustering is detected for all the subsamples of massive galaxies characterized by different stellar masses 
($M_{\star} = 10^{11.0-11.5} M_{\odot}$, $10^{11.5-12.0} M_{\odot}$) or rest-frame colors (blue: $U - V < 1.0$, red: $U - V > 1.0$).
We find that more mature (more massive or redder) galaxies are more clustered, which implies that the more mature galaxies have started 
stellar-mass assembly earlier within the highly-biased region where the structure formation has also started earlier.
By means of halo occupation distribution (HOD) models fitted to the observed angular correlation function, we infer the properties
of the underlying host dark halos.
We find that the estimated bias factors and host halo masses are systematically larger for galaxies with larger stellar masses,
which is consistent with the general agreement that the capability of hosting massive galaxies depends strongly on halo mass.
The estimated effective halo masses are $\sim 10^{14} M_{\odot}$, which gives the stellar-mass to halo-mass ratios of $\sim 0.003$.
The observed evolution of bias factors indicates rapid evolution of spatial distributions of cold dark matter relative to those traced by 
the massive galaxies, while the transition of host halo masses might imply that the fractional mass growth rate of halos is less than those
of stellar systems.
The inferred halo masses and high fractions of central galaxies indicate that the massive galaxies in the
current sample are possibly equivalent to central galaxies of galaxy clusters.
\end{abstract}


\section{Introduction}

Understanding the origin and evolution of galaxies, in particular the most massive, is one of the major challenges in modern astrophysics. 
Many massive galaxies today are giant early-type systems; hence the formation of spheroids should proceed to a certain extent in locked step 
with the mass assembly.
Numerical simulations based on the $\Lambda$ Cold Dark Matter (CDM) theory \citep{white78} predict the hierarchical mass assembly, i.e., large
CDM halos (or dark halos) are formed through successive mergers of smaller building-block systems.
Within formed dark halos, baryons dissipate their energy and gravitationally collapse to form luminous galaxies.
The simplest consequence of the above scenario is that galaxies are also formed in a hierarchical way, which means that 
the most massive galaxies emerge in the last phase of the formation history.
Thus the number-density measurements of massive galaxies, or luminous red galaxies as a proxy of them, in the distant universe are 
one of the key observations in recent studies of cosmology and galaxy evolution.
While there has been growing evidence that luminous red galaxies are largely in place at $z \sim 1$ \citep[e.g.,][]{bell04, borch06, cimatti06, 
willmer06, brown07, faber07},
recently \citet{matsuoka10} found an evidence for the hierarchical formation of massive ($M_{\star} > 10^{11} M_{\odot}$) galaxies since $z = 1$
based on an analysis of a large galaxy sample extracted from the near-infrared images.

Clustering measurements provide another important clue as to the formation and evolution of massive galaxies.
In the structure formation paradigm of the $\Lambda$ CDM theory, dark halos are predicted to cluster together with the clustering strengths
depending on their masses, in such a way that more massive dark halos have larger clustering amplitudes.
Thus measurements of galaxy clustering can be used to infer the underlying CDM distribution, and eventually reveal the evolutionary
path of galaxies throughout the cosmic mass assembly and structure formation history.
There have been a number of clustering measurements of luminous red galaxies in the local \citep[e.g.,][]{norberg02,budavari03,zehavi05} and 
distant \citep[e.g.,][]{brown03, brown08, phleps06, ross07, coil08, mccracken08} universe.
They found that more luminous galaxies are generally more clustered, as expected if more massive (thus more luminous in red wave bands) galaxies 
reside in more massive dark halos.
Clustering measurements of stellar-mass selected galaxies are much more challenging due to the difficulty in estimating stellar masses of a
large number of galaxies over a wide field of sky.
Recently \citet{foucaud10} present such a measurement for galaxies with the stellar masses $M_{\star} > 10^{10} M_{\odot}$ at $0.4 < z < 2.0$,
taken from the Palomar Observatory Wide-field Infrared Survey, and report the evolution of the stellar-mass to total-mass ratios inferred from
the clustering strengths of galaxies.

In this work we present the clustering measurements of massive ($M_{\star} > 10^{11} M_{\odot}$) galaxies at $0.2 < z < 1.0$, 
a sample built up in \citet{matsuoka10}, and show a simple interpretation of the measured angular correlation function 
by using halo occupation distribution (HOD) models.
This paper is organized as follows.
We introduce our galaxy sample and quantify their angular correlation in Section 2.
The construction of HOD models and the comparison with the observations are described in Section 3.
The inferred properties of underlying dark halos are discussed in Section 4, then a summary follows in Section 5.
Throughout this paper, we adopt the cosmological parameters of $H_0$ = $100 h$ = 70 km s$^{-1}$ Mpc$^{-1}$, $\Omega_{\rm M} = 0.3$, $\Omega_{\rm b} = 0.04$,
$\Omega_{\Lambda} = 0.7$, $\sigma_8 = 0.8$, and $n_s = 1.0$.
All magnitudes are expressed in the Vega system.

\section{Measurements}

\subsection{Sample}

The massive galaxy sample used in this work is taken from \citet{matsuoka10}.
Below we give a short description of the sample, while we refer the reader to the above paper for the full details.

The massive galaxies are drawn from 55.2 deg$^2$ of the UKIRT (United Kingdom Infrared Telescope) Infrared Deep Sky 
Survey \citep[UKIDSS;][]{lawrence07} and the Sloan 
Digital Sky Survey \citep[SDSS;][]{york00} II Supernova Survey \citep{frieman08}, which enable by far the largest survey of 
massive galaxies to date with robust mass estimate, reaching back to a time when the universe had only half its present size.
The target field is a strip region from the right ascension 1$^{\rm h}$15$^{\rm m}$ to 3$^{\rm h}$6$^{\rm m}$ along the
celestial equator, i.e., the declination within $\pm 1.25^{\circ}$.
The galaxies have been selected from the $K$-band images of the UKIDSS Large Area Survey Data Release 3 (Warren et al., in prep.), 
and separated from Galactic stars based on their $r-z$ and $z-K$ colours.
We estimate that more than 95 \% of the galaxies in the $K$-band sources are correctly extracted and that stellar contamination
is negligible.
Photometric redshifts have been derived by comparing the galaxy colours in the $u$, $g$, $r$, $i$, $z$, $Y$, $J$, $H$, and $K$ bands
to the spectral energy distribution templates created from a part of the sample with spectroscopic redshifts.
The estimated redshift accuracy is $\sigma_{{\Delta}z/(1+z)} \sim 0.04$.
Stellar masses have been estimated by fitting the stellar population synthesis models of \citet{bc03} to the observed colours, by
varying the age, metallicity, typical star-formation duration, and dust extinction of stellar populations.
We adopted the initial mass function (IMF) of \citet{salpeter55}.
The fairly robust estimates of stellar mass were obtained thanks to the inclusion of near-infrared photometry \citep[e.g.,][]{matsuoka08},
with the estimated uncertainty $\sigma_{\Delta {\rm log} M_{\star}} \sim 0.2$.
The rest-frame $U-V$ colours were calculated by $k$-correcting the observed $r$, $i$, or $z$-band magnitudes.

We define the massive galaxies in this work with the stellar masses $10^{11.0} M_{\odot} < M_{\star} < 10^{12.0} M_{\odot}$.
The galaxies are grouped into four redshift bins, $z$ = 0.2 -- 0.4, 0.4 -- 0.6, 0.6 -- 0.8, and 0.8 -- 1.0.
In addition, we classify the galaxies in the following two ways.
On the one hand, they are divided based on their stellar masses into the $10^{11.0-11.5} M_{\odot}$ and the $10^{11.5-12.0} M_{\odot}$ 
galaxies.
On the other hand, we divide them based on their rest-frame $U - V$ colours, into the blue ($U - V < 1.0$) and the red ($U - V > 1.0$) 
galaxy populations.
The total numbers of the subsamples are summarized in Table \ref{tab:sample_num}.

\begin{table}
 \caption{Total numbers of the massive galaxy subsamples.}
 \label{tab:sample_num}
 \begin{tabular}{@{}ccccc}
  \hline
  Redshift                   & 0.2 -- 0.4 & 0.4 -- 0.6 & 0.6 -- 0.8 & 0.8 -- 1.0\\
  \hline
  $10^{11.0-11.5} M_{\odot}$ & 9720      & 15300     & 18582     & 12371\\
  $10^{11.5-12.0} M_{\odot}$ & 1408      & 572        & 815        & 613\\
  \hline
  blue                       & 1724       & 4402       & 5661       & 4033\\
  red                        & 9404       & 11470      & 13736      & 8951\\
  \hline
 \end{tabular}
\end{table}

While the detection completeness is nearly 100 \% for our sample at $z < 1$, they could have a significant fraction of 
the contamination from less massive galaxies due to the Eddington bias \citep{eddington13}, since they are located at the steep 
high-end portion of galaxy stellar mass function.
It is especially significant for the most massive, $10^{11.5-12.0} M_{\odot}$ galaxies, for which a contaminated fraction
could be up to 50 \% based on the Monte-Carlo simulation.
Such a contamination would generally reduce measured amplitudes of galaxy clustering.

\subsection{Correlation Function}

A common statistic characterizing clustering of galaxies is the angular correlation function (ACF), $w (\theta)$.
It is defined by the probability ${\delta}P$ of finding two galaxies separated by an angular distance $\theta$,
with respect to that expected for a random distribution, i.e.,
\begin{equation}
  {\delta}P \propto S_{\rm gal}^2 [ 1 + w (\theta) ] \delta\Omega_1 \delta\Omega_2 
\end{equation}
where $\delta\Omega_1$ and $\delta\Omega_2$ are elements of solid angle and $S_{\rm gal}$ is the mean surface density of galaxies.
The ACF $w (\theta)$ becomes zero for a random or homogeneous distribution.

We use the \citet{landy93} estimator for evaluating the observed ACFs as follows.
On a field with $N_{\rm G}$ detected galaxies, the numbers of galaxy pairs $N_{\rm GG} (\theta)$ are counted as a function 
of the separation angle $\theta \pm \Delta\theta$ (deg).
Here we adopt the logarithmic step size ${\Delta}$ (log $\theta$) = 0.2. 
On the other hand, $N_{\rm R}$ points are placed at random over the observed fields as covered by the real data (galaxies), and
the numbers of random pairs $N_{\rm RR} (\theta)$ are counted.
In addition, the numbers of data (galaxy)--random pairs, $N_{\rm GR} (\theta)$, are also counted with the above data and random points.
Then the \citet{landy93} estimator gives
\begin{equation}
  w^{*} (\theta) = \frac{DD(\theta) - 2 DR(\theta) + RR(\theta)}{RR(\theta)},
\end{equation}
where $DD$, $DR$, and $RR$ are the data--data, data--random, and random--random pair counts normalized by the total numbers
of pairs, i.e., $DD = N_{\rm GG}/[\frac{1}{2} N_{\rm G} (N_{\rm G}-1)]$, 
$DR = N_{\rm GR}/[N_{\rm G} N_{\rm R}]$, and $RR = N_{\rm RR}/[\frac{1}{2} N_{\rm R} (N_{\rm R}-1)]$.
We repeat the calculation 10 times with varying the random components.

The measured $w^* (\theta)$ are known to be biased to lower values with respect to the real $w (\theta)$ due to 
the boundary effect of the observed area.
This bias offset is known as the integral constraint, $w_{\rm IC}$ \citep{groth77}.
We estimate its amplitude by approximating the ACF with the power-law form 
\begin{equation}
w (\theta) = w^* (\theta) + w_{\rm IC} = A_w \theta^{-\delta} .
\end{equation}
In the case of a power-law ACF, the integral constraint is given from the random--random pair counts by 
$w_{\rm IC}/A_w = \sum_{\theta} N_{\rm RR} (\theta) \theta^{-\delta} / \sum_{\theta} N_{\rm RR} (\theta)$ 
\citep[e.g.,][]{roche99}.
We assume the standard value of the ACF slope $\delta = 0.8$, which is found in a wide range of optical and infrared observations 
of nearby and distant galaxies \citep[e.g.,][]{baugh96, roche99b}.
While the steeper slopes ($\delta \sim 1$) have been reported for nearby early-type galaxies \citep[e.g., ][]{loveday95, guzzo97}
and luminous red galaxies \citep[e.g.,][]{brown08}, we confirmed that it is sufficient to adopt a single value of $\delta = 0.8$ 
for our sample considering the relatively large errors in the ACF measurements especially for the most massive galaxies.
The estimated amplitudes of the integral constraint are $w_{\rm IC} \sim 0.04\ w (1')$, which are negligible at most of the angular 
scales studied in this work.

\begin{figure*}
  \includegraphics[width=168mm]{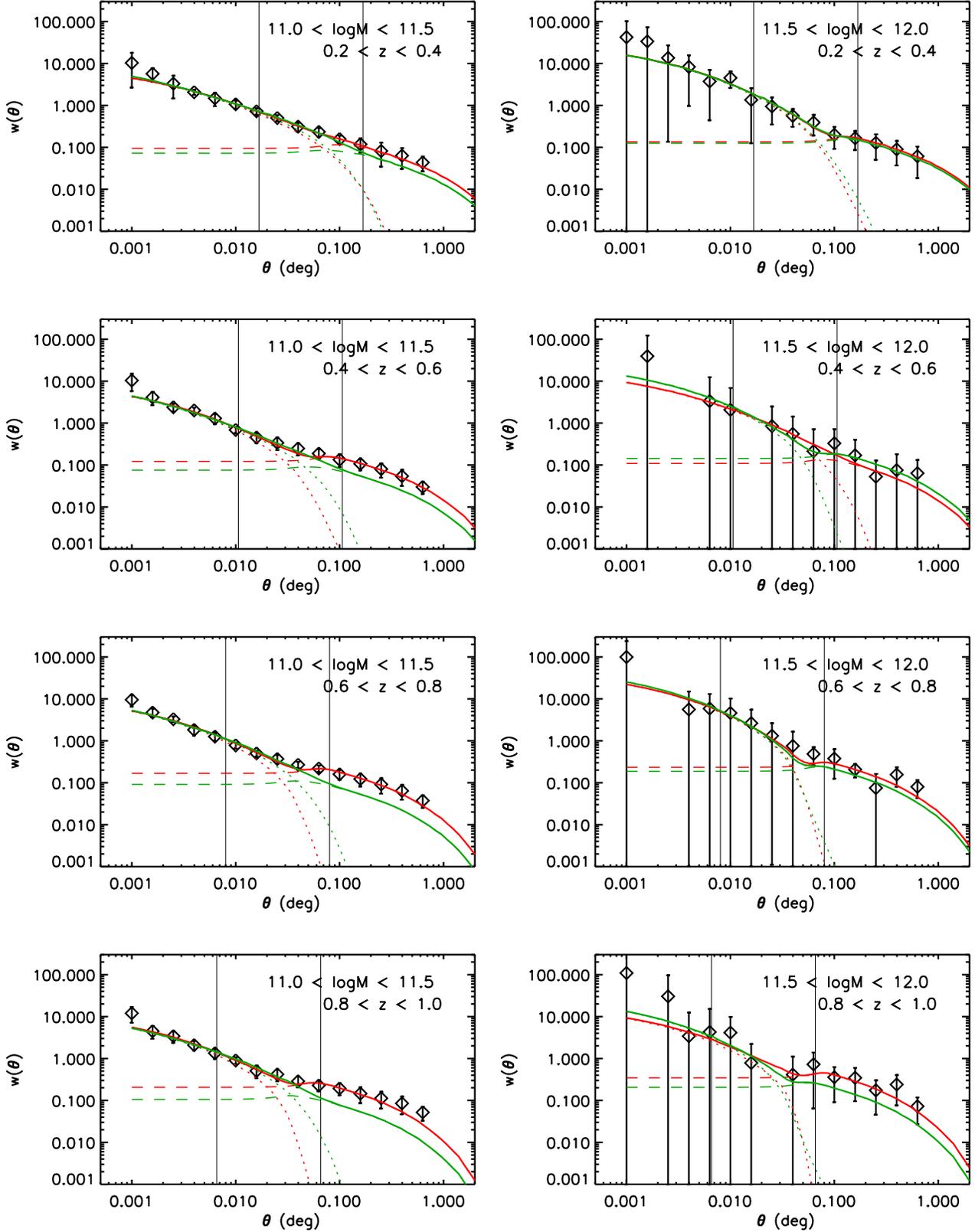}
  \caption{The measured angular correlation functions (diamonds) for the $10^{11.0-11.5} M_{\odot}$ (left panels) and 
    the $10^{11.5-12.0} M_{\odot}$ (right panels) galaxies at $z$ = 0.2 -- 0.4, 0.4 -- 0.6, 0.6 -- 0.8, and 0.8 -- 1.0,
    from top to bottom.
    The error bars represent the 2$\sigma$ uncertainties of the measurements.
    The green and red curves show the best-fit HOD models of the case (i) and (ii), respectively (see text); the dotted and dashed lines 
    represent the one-halo and two-halo terms, respectively, and the solid lines represent their sums.
    The thin vertical lines mark the comoving scales of 0.5 and 5 $h^{-1}$ Mpc at each redshift.}
  \label{acf_results}
\end{figure*}

\begin{figure*}
  \includegraphics[width=168mm]{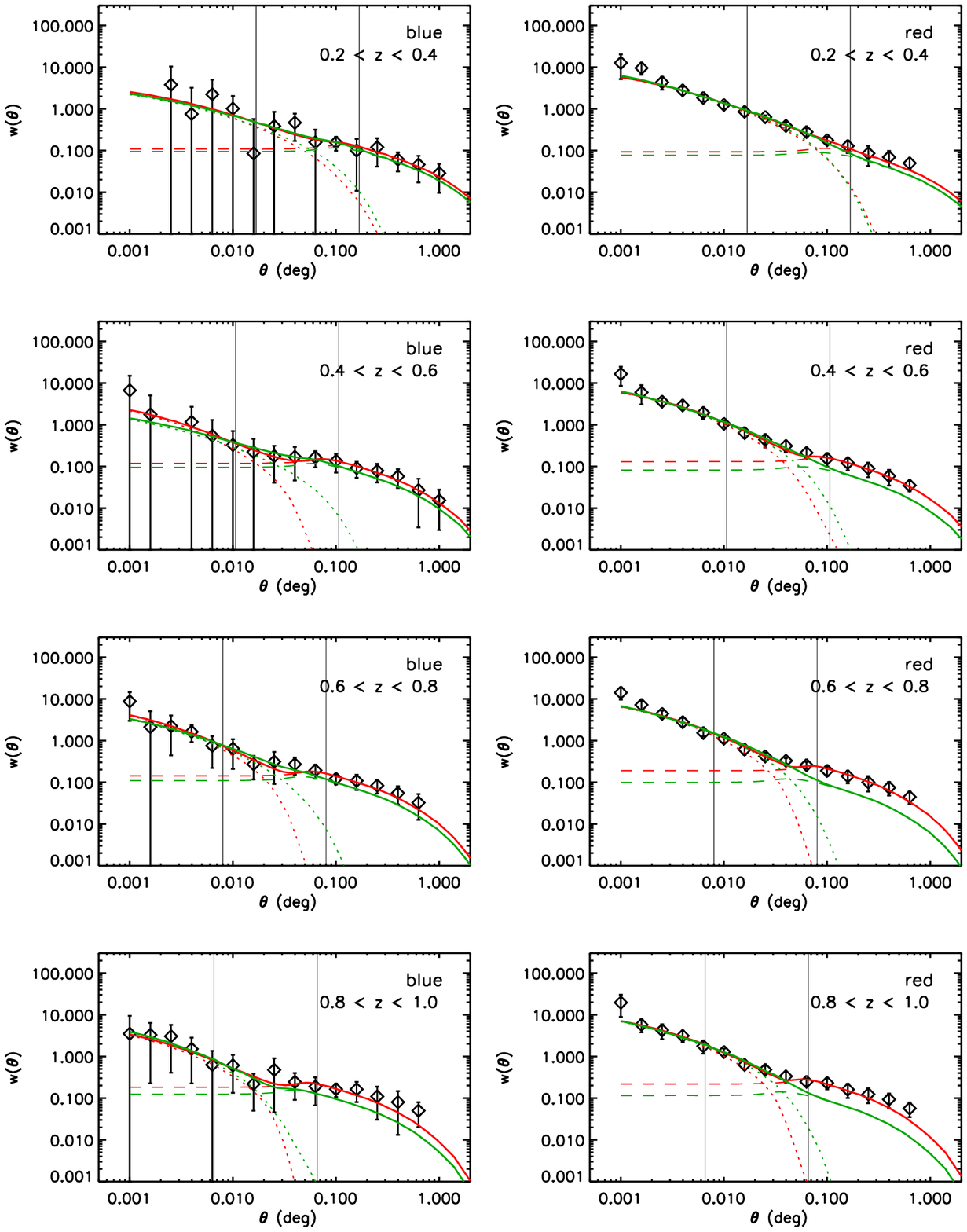}
  \caption{Same as Fig. \ref{acf_results}, but for the blue (left panels) and the red (right panels) galaxy populations.}
  \label{acf_results_color}
\end{figure*}

We divide our strip-shaped field into nine sub-fields along the right ascension, each covering approximately 
$3.0 \times 2.5$ deg$^2$, and measure the ACF in each of the sub-fields.
The measurements in the sub-fields are then averaged to give the mean estimates of $w (\theta)$ with the associated
errors calculated from the field-to-field scatters by the bootstrap resampling.
The results are shown in Figs. \ref{acf_results} and \ref{acf_results_color} for the different stellar-mass, color, and
redshift classes.
Note that one should take the covariance matrices into account in rigorous analyses of the ACFs, since the measured ACF amplitudes
at different angular scales are correlated with each other.
However, it would have little impact on the conclusions presented in this work, given that the uncertainties of the deduced halo 
characteristics come mostly from the inadequacy of the current HOD models rather than the measurement errors (see below).

\section{Halo Occupation Distribution Models\label{sec:hod2}}

We investigate the evolutionary link of the observed galaxies to underlying dark (CDM) halos by using halo occupation distribution 
\citep[HOD; see][for a review]{cooray02} models.
HOD modeling is a powerful approach to explore galaxy distribution within dark halos, whose formation and evolution can be predicted by
simulations and analytic methods.
Observations can provide the useful insights into underlying dark halos by constraining HOD parameters in such a way that 
the models appropriately reproduce measured clustering properties of galaxies \citep[e.g.,][]{zehavi05, lee06}.
In this work we follow \citet{blake08} for constructing our HOD models \citep[see also][]{tinker05}.
We give a short description of the models below, while the reader is referred to the above papers and the references therein for further details.

\subsection{Model Construction \label{sec:hod}}

In the HOD models we use the halo mass function $n (M)$ presented by \citet{jenkins01}.
We assume the scale-dependent bias following \citet{tinker05}, with the bias function $b (M)$ taken from the \citet{sheth01} model including the 
revised parameters given by \citet{tinker05}.
The non-linear dark-matter power spectrum is constructed using the fitting formulae of \citet{smith03}.
The dark halo density profile of \citet{nfw97} is adopted, with the characterizing concentration parameter $c (M)$ calibrated by 
numerical simulations \citep{bullock01}.

\begin{figure}
  \includegraphics[width=84mm]{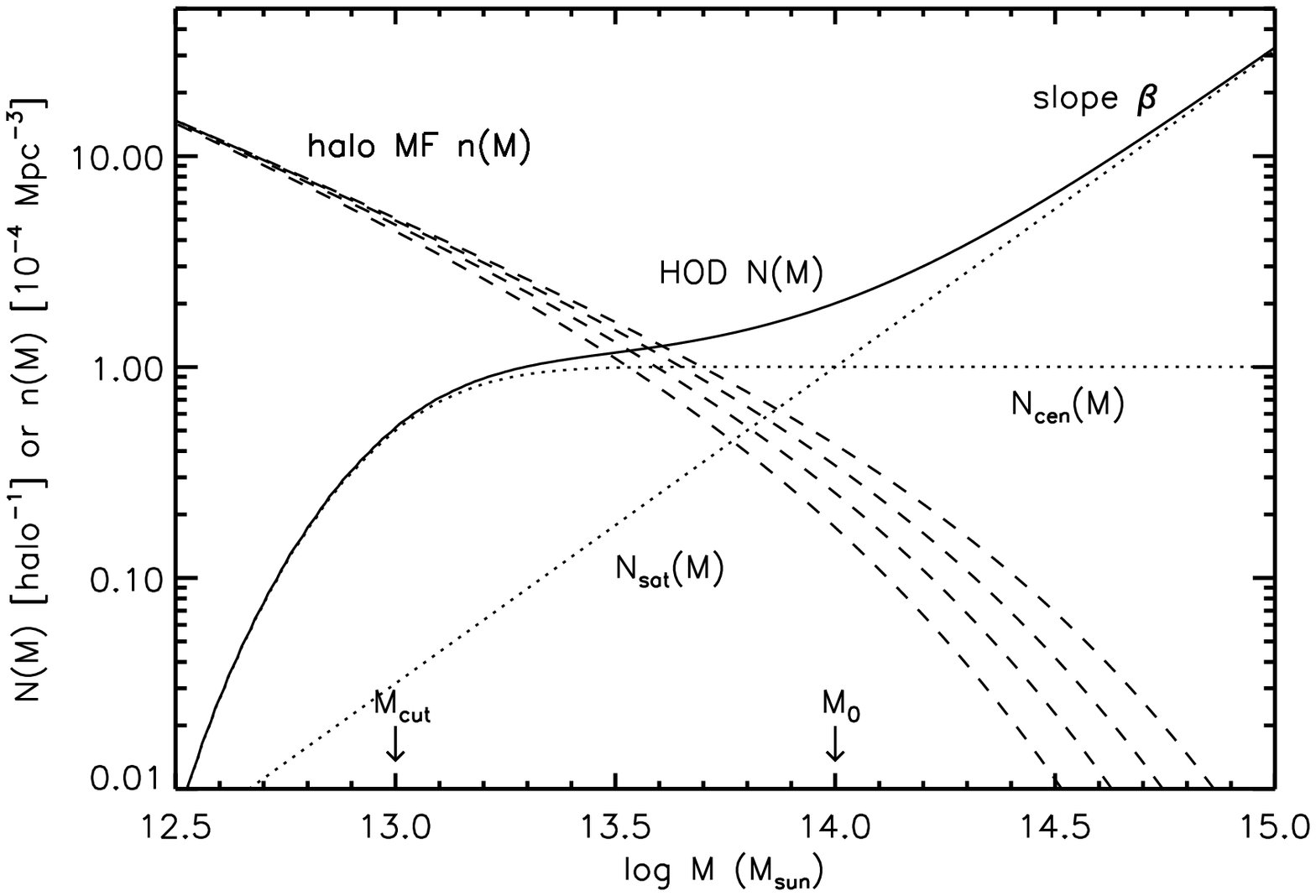}
  \caption{An example of the HOD model $N(M)$ and the halo mass function $n(M)$.
    The HOD model is calculated with the parameter values $M_{\rm cut} = 10^{13} M_{\odot}$, $M_0 = 10^{14} M_{\odot}$, and $\beta = 1.5$.
    The dotted lines represent the numbers of central galaxies $N_{\rm cen}(M)$ and satellite galaxies $N_{\rm sat}(M)$, and the solid
    lines represent the total number $N (M) = N_{\rm cen} (M) [1 + N_{\rm sat} (M)]$.
    The dashed lines show the evolving halo mass function at $z$ = 0.3, 0.5, 0.7, and 0.9, from top to bottom.
    An actual number of galaxies hosted in a halo with the mass $M$ in a given volume scales as $n (M) \times N(M)$.}
  \label{hod_ex}
\end{figure}

The statistical number of a central galaxy within a dark halo of the mass $M$ is parametrized as
\begin{equation}
  N_{\rm cen} (M) = 0.5 \left[ 1 + {\rm erf} \left( \frac{{\rm log}\ (M/M_{\rm cut})}{\sigma_{\rm cut}} \right) \right] ,
  \label{eq:hod_c}
\end{equation}
where erf ($x$) is the error function \citep{zheng05}.
The halos above a threshold mass of $M_{\rm cut}$ have a single central galaxy and those below $M_{\rm cut}$ have no central galaxies on average,
while the transition between the two cases is soften by the parameter $\sigma_{\rm cut}$. 
We adopt $\sigma_{\rm cut} = 0.3$, which is the mean value found for luminous red galaxies at $z < 1$ \citep{blake08, brown08}.
The number of satellite galaxies is described as the power-law form
\begin{equation}
  N_{\rm sat} (M) = \left( \frac{M}{M_0} \right)^{\beta} .
  \label{eq:hod_s}
\end{equation}
The total number of galaxies populating a dark halo is given by 
\begin{equation}
  N (M) = N_{\rm cen} (M) [1 + N_{\rm sat} (M)] , 
\end{equation}
thus no galaxies are statistically found in the dark halos below the threshold mass $M_{\rm cut}$.
We show an example of our HOD model along with the adopted halo mass function in Fig. \ref{hod_ex}.

\begin{figure}
  \includegraphics[width=84mm]{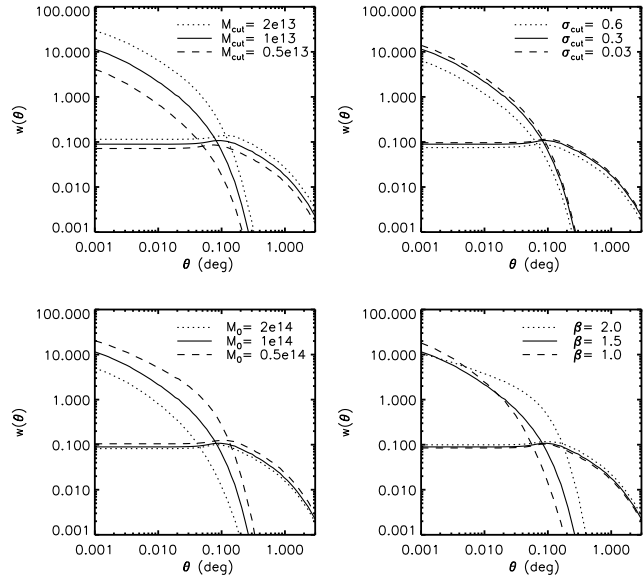}
  \caption{The dependence of the predicted ACF on the HOD parameters.
    In each panel the two curves of the same line style represent the one-halo and the two-halo terms, the former being dominant at small angular scales.
    The basic parameters are ($M_{\rm cut}$, $\sigma_{\rm cut}$, $M_0$, $\beta$) = ($10^{13} M_{\odot}$, 0.3, $10^{14} M_{\odot}$, 1.5) and one of the
    parameter values is altered in each panel as indicated at the top right corner.
    Note that $\sigma_{\rm cut}$ is fixed to 0.3 in the present work.}
  \label{blake08_hod_para}
\end{figure}

A spatial clustering of galaxies $\xi (r)$ can be divided into the two terms, namely, a one-halo term ($\xi_{\rm 1h} (r)$) arising from pairs 
of galaxies in a same dark halo, and a two-halo term ($\xi_{\rm 2h} (r)$) arising from pairs of galaxies that reside in different halos:
\begin{equation}
\xi (r) = [1 + \xi_{\rm 1h} (r)] + \xi_{\rm 2h} (r) .
\end{equation}
The one-halo term dominates a correlation function at small scales $\la$ a few $h^{-1}$ Mpc, i.e., less than a typical virial diameter of large halos,  
while a two-halo term becomes dominant at larger scales.
The halo exclusion (which implements the condition that halos cannot reside within each other) is considered by introducing the truncation mass, 
above which the halos would overlap with each other at the separation $r$, into the calculation of $\xi_{\rm 2h} (r)$.
The truncation mass is derived using the '$n'_g$-matched' approximation described in \citet{tinker05}.
The spatial correlation $\xi (r)$ is projected to the angular correlation function $w (\theta)$ via
\begin{equation}
w (\theta) = 2 \int_0^{\infty} dx\ \left[ \frac{p (z)}{dx (z)/dz} \right]^2 \int_0^{\infty} du\ \xi (\sqrt{u^2 + x (z)^2 \theta^2}) ,
\end{equation}
where $x (z)$ is the comoving radial distance to the redshift $z$.
The redshift distribution of galaxies, $p (z)$, is estimated by giving the random gaussian errors with the standard deviation
$\sigma_{\Delta z/(1+z)} \sim 0.04$, which corresponds to our photometric-redshift uncertainty, to the interpolated number-density evolution 
reported in \citet{matsuoka10} (including the local measurements by \citet{cole01}) multiplied by the comoving volume element, and
then extracting the true redshift distribution of the sources found in the disturbed-redshift bin under consideration.
However, we find that the resultant HOD parameters presented below are not very sensitive to the details of the $p(z)$ function.
We show the examples of the calculated ACFs with the several sets of the HOD parameters in Fig. \ref{blake08_hod_para}.

From the above parametrization of the HOD $N (M)$ along with the halo mass function $n (M)$ and the bias function $b (M)$, we can calculate some 
useful quantities; the number density of galaxies:
\begin{equation}
  n_{\rm gal} = \int dM\ n(M)\ N(M) ,
\end{equation}
the effective large-scale bias:
\begin{equation}
  b_{\rm gal} = \int dM\ b(M)\ n(M)\ \frac{N(M)}{n_{\rm gal}} ,
\end{equation}
the effective mass of the host dark halo:
\begin{equation}
  M_{\rm eff} = \int dM\ M\ n(M)\ \frac{N(M)}{n_{\rm gal}} ,
  \label{eq:meff}
\end{equation}
and the average fraction of central or satellite galaxies:
\begin{equation}
  \begin{split}
    f_{\rm cen} = & \frac{1}{n_{\rm gal}} \int dM\ n(M)\ N_{\rm cen}(M);\ {\rm central} ,\\
    f_{\rm sat} = & 1 - f_{\rm cen};\ {\rm satellite} .
    \end{split}
\end{equation}

\subsection{Other Models \label{sec:otherhod}}

Note that the above HOD form is based on a theoretical galaxy sample above a certain baryonic mass threshold, predicted by smoothed particle hydrodynamics 
(SPH) simulations and semi-analytic galaxy formation models \citep{berlind03, zheng05}.
While our galaxy sample is defined by the stellar-mass bins rather than thresholds, we use the above models in this work for the following reasons.
First, our massive galaxies are located at the steep high-end part of the mass function, so that the galaxies above the upper limit of a stellar-mass 
bin have little impact on the observables whether they are included in the measurements or not.
In fact we find no significant difference in the measured number densities and ACFs between the $M_{\star} = 10^{11.0-11.5} M_{\odot}$ and
$M_{\star} > 10^{11.0} M_{\odot}$ galaxy samples or between the $M_{\star} = 10^{11.5-12.0} M_{\odot}$ and $M_{\star} > 10^{11.5} M_{\odot}$ galaxy samples.
Second, we actually find that considering an upper cut-off in $N_{\rm cen} (M)$, which is used to reproduce the HOD of a bin sample (see below), 
does not provide the better fits to the observed ACFs than our standard model does. 
Third, the adopted model can approximate the HODs of various kinds of galaxies with a minimum number of parameters \citep{zheng05}, which is essential
in exploring the halo characteristics of galaxies with relatively large errors in the ACF measurements, such as our sample.

We have examined the effects of changing HOD models as follows.
The standard model is modified by considering the variable $\sigma_{\rm cut}$, fixed $\beta$ and/or the additional cut-off parameter $M_1$ of the 
satellite HOD:
\begin{equation}
  N_{\rm sat} (M) = \left( \frac{M - M_1}{M_0} \right)^{\beta} .
  \label{eq:hod_s2}
\end{equation}
We have also tried the six-parameter form:
\begin{equation}
  \begin{split}
    N_{\rm cen} (M) = &\ N_0\ {\rm exp} \left[ - \frac{{{\rm log}\ (M/M_{\rm cen})}^2}{2 \sigma^2} \right]\ \ (N_0 < 1)\,\\
    N_{\rm sat} (M) = &\ \left( \frac{M - M_1}{M_0} \right) ^{\beta} ,
  \end{split}
  \label{eq:hodsix}
\end{equation}
which is constructed to reproduce the HODs of a sample of galaxies in a stellar-mass bin or of young (in terms of stellar age) galaxies presented by \citet{zheng05}.
The essential difference of this form from our standard one is that it allows an upper cut-off of $N_{\rm cen} (M)$.
In addition, we have replaced the whole model construction process (from the choice of dark-matter power spectrum) from the above one following \citet{blake08} 
to the one following the recipe given by \citet{hamana04}, who adopt the HOD form of
\begin{equation}
  N (M) = \begin{cases}
                      (M/M_1)^{\alpha} & {\rm for}\ M > M_{\rm min} ,\\
                      0                & {\rm for}\ M < M_{\rm min} .
                    \end{cases}
\end{equation}
We have also tried a few other halo mass functions \citep{sheth99, reed07, bhattacharya10} and bias functions \citep{sheth99, tinker10b}.
However, we find that none of the above altered models provide the better reproduction of the observed ACFs than our standard model does.

\subsection{Results \label{sec:hodresults}}

We constrain the parameters $M_{\rm cut}$, $M_0$, and $\beta$ of the standard HOD model by comparing the predicted ACFs ($w^{\rm th} (\theta)$) 
and number densities ($n_{\rm gal}^{\rm th}$) with those observed ($w^{\rm obs} (\theta)$ and $n_{\rm gal}^{\rm obs}$) through the Markov chain 
Monte Carlo (MCMC) method.
The observed number densities are taken from \citet{matsuoka10}. 
We evaluate the likelihood of the specific model by the $\chi^2$ value of the model fit to the observed quantities.
The ACF fitting is restricted to the angular scales $\theta < $ 0.3 deg where the effects of the integral constraint (i.e., the boundary
effect of the observations) are negligible.

We summarize the obtained best-fit HOD parameters in Table \ref{tab:hod} (the former values in the parentheses).
In the case of the $10^{11.5 - 12.0} M_{\odot}$ galaxies at $0.6 < z < 0.8$, the parameter $\beta$ is not well constrained in the plausible range
($\beta < 3$) and we fix it to the best-fit value found for the $10^{11.0 - 11.5} M_{\odot}$ galaxies at the same redshift range
(it is justified by the fact that we find the consistent $\beta$ values for the two populations at other redshifts).
The poor constraint on $\beta$ is caused by the fact that the three HOD parameters correlate with each other in reproducing a given ACF, hence 
the MCMC fitting could converge to unrealistic combinations of the parameter values when a measured ACF is accompanied by large uncertainty.
In other subsamples, plausible values of $\beta$ are obtained; we find $\beta = 1.5 - 2.0$ for all but the blue subsamples, which
are in good agreement with those found for SDSS luminous red galaxies at $0.4 < z < 0.7$ measured by \citet{blake08}.

The ACFs of the best-fit HOD models are overlaid on the observed ACFs in Figs. \ref{acf_results} and \ref{acf_results_color} (green curves).
In these fittings, the systematic discrepancy between the model and measured ACFs are observed at the large angular scales where two-halo terms dominate
the ACFs. 
The reason for this discrepancy is fairly clear; the observed ACFs are so strong that very massive halos are required to reproduce them, while the predicted 
number of such massive halos is very small compared to the observed numbers of massive galaxies (the further discussion is given below). 
As a reference, we have carried out another set of model fittings with the modified requirement that the models should only reproduce the observed ACFs,
imposing no constraint on the predicted number densities of galaxies.
We refer to this second set of model fittings as case (ii), and the original fittings as case (i) hereafter.
The results of the former case are listed in Table \ref{tab:hod} (the latter values in the parentheses) and are shown in Figs. \ref{acf_results} and 
\ref{acf_results_color} by the red curves.
We obtain the excellent fits of the ACFs in the case (ii) over the entire angular scales, while the predicted $n_{\rm gal}^{\rm th}$ are smaller than 
$n_{\rm gal}^{\rm obs}$ in the most cases.



\begin{table*}
\rotatebox{90}{
  \begin{minipage}{\textheight}
  \begin{center}
    \caption{Halo Occupation Distributions of Massive Galaxies.$^{1, 2}$ \label{tab:hod}}
    \begin{tabular}{@{}clccccccccc}
      \hline\hline
       Redshift  & Galaxy    &                        & $n_{\rm gal}^{\rm obs}$     &                   &           &         &               &                   &               & \\
         $z$     & sample    &  $<$log $M_{\star}$$>$ & ($10^{-4}\ {\rm Mpc}^{-3}$) & log $M_{\rm cut}$ & log $M_0$ & $\beta$ & $b_{\rm gal}$ & log $M_{\rm eff}$ & $f_{\rm cen}$ & $n_{\rm gal}^{\rm th}$ / $n_{\rm gal}^{\rm obs}$ \\
  \hline
     0.2 -- 0.4  & $10^{11.0-11.5} M_{\odot}$ & 11.2 & 5.5 $\pm$ 0.4   & (12.7, 13.4) & (13.9, 14.6) & (1.6, 2.0) &  (1.53, 1.86) & (13.8, 13.9) & (0.83, 0.94) & (0.97, 0.18)\\
                 &                &   &   & $\pm$ (0.1, 0.4) & $\pm$ (0.1, 0.3) & $\pm$ (0.1, 0.1) & $\pm$ (0.01, 0.12) & $\pm$ (0.1, 0.1) & $\pm$ (0.01, 0.02) & $\pm$ (0.06, 0.09)\\
                 & $10^{11.5-12.0} M_{\odot}$ & 11.7 & 0.49 $\pm$ 0.08 & (13.7, 13.8) & (14.7, 15.0) & (1.6, 1.4) &  (2.13, 2.26) & (14.0, 14.1) & (0.92, 0.95) & (0.98, 0.65)\\
                 &                &   &   & $\pm$ (0.1, 0.4) & $\pm$ (0.2, 0.6) & $\pm$ (0.2, 0.2) & $\pm$ (0.05, 0.18) & $\pm$ (0.1, 0.2) & $\pm$ (0.01, 0.03) & $\pm$ (0.14, 0.39)\\
                 & blue                       & 11.2 & 0.9 $\pm$ 0.4   & (13.4, 13.7) & (14.7, 14.9) & (2.4, 2.4) &  (1.90, 2.10) & (13.9, 14.0) & (0.97, 0.98) & (0.88, 0.48)\\
                 &                &   &   & $\pm$ (0.4, 0.5) & $\pm$ (0.4, 1.2) & $\pm$ (0.2, 0.4) & $\pm$ (0.15, 0.23) & $\pm$ (0.2, 0.3) & $\pm$ (0.01, 0.01) & $\pm$ (0.39, 0.32)\\
                 & red                        & 11.3 & 5.0 $\pm$ 0.5   & (12.8, 13.3) & (13.9, 14.5) & (1.5, 1.9) &  (1.58, 1.83) & (13.8, 13.9) & (0.81, 0.92) & (0.95, 0.25)\\
                 &                &   &   & $\pm$ (0.1, 0.3) & $\pm$ (0.1, 0.3) & $\pm$ (0.1, 0.1) & $\pm$ (0.01, 0.09) & $\pm$ (0.1, 0.1) & $\pm$ (0.01, 0.02) & $\pm$ (0.09, 0.10)\\
  \hline
     0.4 -- 0.6  & $10^{11.0-11.5} M_{\odot}$ & 11.2 & 3.9 $\pm$ 0.3   & (12.9, 13.7) & (14.0, 15.0) & (1.6, 1.8) &  (1.73, 2.46) & (13.7, 14.0) & (0.86, 0.98) & (0.93, 0.07)\\
                 &                &   &   & $\pm$ (0.1, 0.2) & $\pm$ (0.1, 0.3) & $\pm$ (0.1, 0.1) & $\pm$ (0.02, 0.10) & $\pm$ (0.1, 0.1) & $\pm$ (0.01, 0.01) & $\pm$ (0.07, 0.02)\\
                 & $10^{11.5-12.0} M_{\odot}$ & 11.6 & 0.10 $\pm$ 0.02 & (14.0, 13.6) & (15.2, 14.6) & (1.6, 2.2) &  (2.97, 2.45) & (14.2, 14.0) & (0.97, 0.92) & (1.03, 4.40)\\
                 &                &   &   & $\pm$ (0.1, 0.6) & $\pm$ (0.6, 0.7) & $\pm$ (0.2, 0.5) & $\pm$ (0.07, 0.29) & $\pm$ (0.1, 0.3) & $\pm$ (0.02, 0.09) & $\pm$ (0.15, 4.69)\\
                 & blue                       & 11.2 & 1.1 $\pm$ 0.2   & (13.3, 13.7) & (14.6, 16.3) & (2.4, 0.9) &  (1.98, 2.33) & (13.8, 13.9) & (0.98, 0.99) & (0.89, 0.32)\\
                 &                &   &   & $\pm$ (0.1, 0.2) & $\pm$ (0.1, 0.8) & $\pm$ (0.3, 0.4) & $\pm$ (0.05, 0.12) & $\pm$ (0.1, 0.2) & $\pm$ (0.01, 0.01) & $\pm$ (0.14, 0.10)\\
                 & red                        & 11.2 & 2.9 $\pm$ 0.3   & (13.0, 13.8) & (14.0, 15.0) & (1.6, 2.0) &  (1.86, 2.61) & (13.8, 14.1) & (0.83, 0.97) & (0.93, 0.07)\\
                 &                &   &   & $\pm$ (0.1, 0.2) & $\pm$ (0.1, 0.5) & $\pm$ (0.1, 0.3) & $\pm$ (0.02, 0.12) & $\pm$ (0.1, 0.1) & $\pm$ (0.01, 0.01) & $\pm$ (0.08, 0.02)\\
  \hline
     0.6 -- 0.8  & $10^{11.0-11.5} M_{\odot}$ & 11.2 & 3.4 $\pm$ 0.3   & (12.9, 13.9) & (13.9, 15.2) & (1.7, 1.7) &  (1.96, 3.04) & (13.6, 14.0) & (0.85, 0.99) & (0.91, 0.04)\\
                 &                &   &   & $\pm$ (0.1, 0.1) & $\pm$ (0.1, 0.3) & $\pm$ (0.1, 0.2) & $\pm$ (0.01, 0.08) & $\pm$ (0.1, 0.1) & $\pm$ (0.01, 0.01) & $\pm$ (0.07, 0.01)\\
                 & $10^{11.5-12.0} M_{\odot}$ & 11.6 & 0.09 $\pm$ 0.02 & (14.0, 14.3) & (15.0, 15.5) & (1.7, 1.6) &  (3.34, 3.90) & (14.2, 14.3) & (0.96, 0.99) & (0.88, 0.27)\\
                 &                &   &   & $\pm$ (0.1, 0.2) & $\pm$ (0.2, 0.5) & $\pm$ (0.0, 0.0) & $\pm$ (0.09, 0.23) & $\pm$ (0.1, 0.2) & $\pm$ (0.01, 0.01) & $\pm$ (0.18, 0.13)\\
                 & blue                       & 11.2 & 1.0 $\pm$ 0.1   & (13.3, 13.7) & (14.4, 15.6) & (2.2, 1.1) &  (2.24, 2.70) & (13.7, 13.9) & (0.95, 0.99) & (0.89, 0.25)\\
                 &                &   &   & $\pm$ (0.1, 0.2) & $\pm$ (0.1, 0.6) & $\pm$ (0.2, 0.4) & $\pm$ (0.04, 0.10) & $\pm$ (0.1, 0.1) & $\pm$ (0.01, 0.01) & $\pm$ (0.10, 0.06)\\
                 & red                        & 11.2 & 2.5 $\pm$ 0.2   & (13.0, 14.0) & (14.0, 15.3) & (1.8, 1.9) &  (2.10, 3.32) & (13.7, 14.1) & (0.85, 0.99) & (0.90, 0.03)\\
                 &                &   &   & $\pm$ (0.1, 0.1) & $\pm$ (0.1, 0.3) & $\pm$ (0.1, 0.2) & $\pm$ (0.02, 0.10) & $\pm$ (0.1, 0.1) & $\pm$ (0.01, 0.01) & $\pm$ (0.08, 0.01)\\
  \hline 
     0.8 -- 1.0  & $10^{11.0-11.5} M_{\odot}$ & 11.2 & 1.9 $\pm$ 0.2   & (13.1, 14.0) & (14.0, 15.4) & (2.0, 1.7) &  (2.31, 3.71) & (13.6, 14.1) & (0.89, 0.99) & (0.92, 0.03)\\
                 &                &   &   & $\pm$ (0.1, 0.1) & $\pm$ (0.1, 0.4) & $\pm$ (0.1, 0.2) & $\pm$ (0.02, 0.12) & $\pm$ (0.1, 0.1) & $\pm$ (0.01, 0.01) & $\pm$ (0.08, 0.01)\\
                 & $10^{11.5-12.0} M_{\odot}$ & 11.6 & 0.06 $\pm$ 0.01 & (14.0, 14.6) & (15.0, 16.4) & (2.1, 1.6) &  (3.72, 5.34) & (14.1, 14.5) & (0.98, 1.00) & (0.93, 0.05)\\
                 &                &   &   & $\pm$ (0.1, 0.3) & $\pm$ (0.5, 0.7) & $\pm$ (0.3, 0.3) & $\pm$ (0.10, 0.45) & $\pm$ (0.1, 0.2) & $\pm$ (0.01, 0.01) & $\pm$ (0.19, 0.04)\\
                 & blue                       & 11.2 & 0.59 $\pm$ 0.06 & (13.4, 13.9) & (14.6, 16.9) & (1.8, 0.8) &  (2.58, 3.42) & (13.7, 14.0) & (0.97, 1.00) & (0.94, 0.15)\\
                 &                &   &   & $\pm$ (0.1, 0.2) & $\pm$ (0.6, 0.9) & $\pm$ (0.3, 0.5) & $\pm$ (0.05, 0.18) & $\pm$ (0.1, 0.1) & $\pm$ (0.01, 0.01) & $\pm$ (0.10, 0.05)\\
                 & red                        & 11.2 & 1.3 $\pm$ 0.1   & (13.2, 14.1) & (14.1, 15.2) & (2.0, 2.3) &  (2.46, 3.87) & (13.7, 14.1) & (0.89, 0.99) & (0.89, 0.03)\\
                 &                &   &   & $\pm$ (0.1, 0.1) & $\pm$ (0.1, 0.5) & $\pm$ (0.1, 0.3) & $\pm$ (0.02, 0.13) & $\pm$ (0.1, 0.1) & $\pm$ (0.01, 0.01) & $\pm$ (0.06, 0.01)\\
   \hline
    \end{tabular}
  \end{center}
  \centering{Note ($^1$) --- The two values in the parentheses represent the best-fit HOD parameters in the case (i) (former) and (ii) (latter).}\\
  \centering{Note ($^2$) --- All masses are given in units of $M_{\odot}$.}
  \end{minipage}
}
\end{table*}

\section{Discussion}

\begin{figure*}
  \includegraphics[width=144mm]{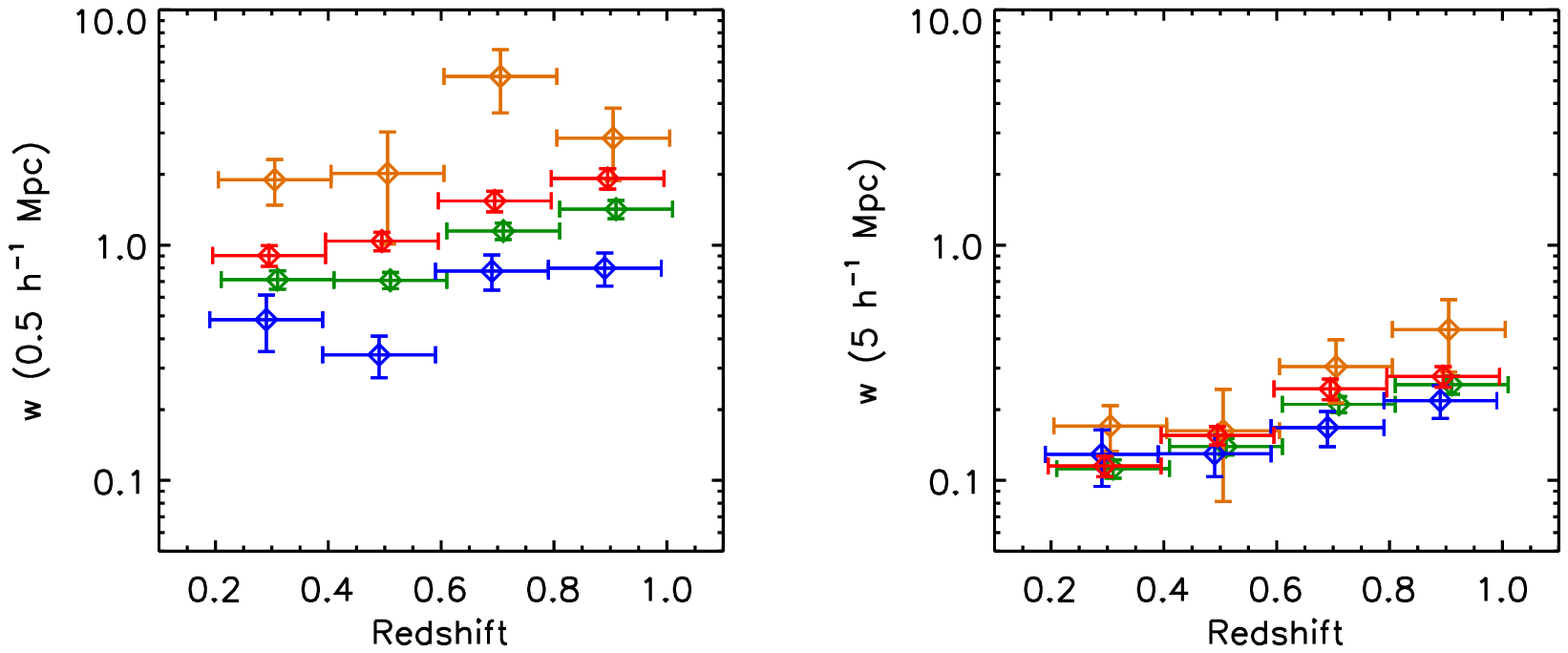}
  \caption{The ACF amplitudes at the comoving scales of $0.5\ h^{-1}$ Mpc (left) and $5\ h^{-1}$ Mpc (right) measured for 
    the $10^{11.0-11.5} M_{\odot}$ (green), the $10^{11.5-12.0} M_{\odot}$ (orange), the blue (blue), and the red (red) galaxies.
    The data points are slightly shifted along the redshift axis relative to each other for visibility.}
  \label{Aw}
\end{figure*}

As a starting point for the discussion, we show the amplitudes of the measured ACFs at the comoving scales of 0.5 and 5 $h^{-1}$ Mpc in Fig. \ref{Aw}.
They are derived by interpolating the data points in Figs. \ref{acf_results} and \ref{acf_results_color} with the best-fit HOD models of the 
case (ii).\footnote{
This does not mean that we accept the halo characteristics behind the best-fit HOD models of the case (ii).
We just use these models as the useful templates to interpolate between the measurement points in Figs. \ref{acf_results} and \ref{acf_results_color},
taking advantage of the fact that the models provide the excellent fits to the measured ACFs.}
The two comoving scales are chosen to represent the clustering within a halo (i.e., one-halo term) or between halos 
(i.e., two-halo term), considering the typical diameter of a few $h^{-1}$ Mpc for large halos.
Apparently the ACF amplitudes are larger in more massive or redder galaxies at both scales.
In other words, we find that more mature (hence massive or red) galaxy systems are more clustered.
It is the expected result if the more mature galaxies have started stellar-mass assembly earlier within the highly-biased region where 
the structure formation has also started earlier.
The larger diversity of the ACF amplitudes at 0.5 $h^{-1}$ Mpc than at 5 $h^{-1}$ Mpc reflects the fact that the one-halo term is much more dependent
on the HOD details of each galaxy population than the two-halo term is.
The two-halo term is mainly determined by the spatial distribution of host halos rather than those of galaxies within a halo, hence its dependence on
the HOD parameters are much weaker than that of the one-halo term.
This is clearly seen in Fig. \ref{blake08_hod_para}.

\begin{table}
 \caption{Two cases of the HOD model fitting.}
 \label{tab:hodcases}
 \begin{tabular}{@{}lcc}
  \hline
  Case  & Fitting results  & Bias or Halo mass\\
  \hline
  (i)   & $w^{\rm th} (\theta) < w^{\rm obs} (\theta)$, $n_{\rm gal}^{\rm th} = n_{\rm gal}^{\rm obs}$ & lower limit \\
  (ii)  & $w^{\rm th} (\theta) = w^{\rm obs} (\theta)$, $n_{\rm gal}^{\rm th} < n_{\rm gal}^{\rm obs}$ & upper limit \\
  \hline
 \end{tabular}
 
Note  --- See text for the details.\\
\end{table}

In the HOD model fitting, we consider the two cases with the different requirements for the models.
They are defined in \S \ref{sec:hodresults} and are summarized in Table \ref{tab:hodcases}.
As already noted, the observed ACFs are so strong that very massive halos are required to reproduce them, while such massive halos are extremely 
rare and could not host the observed number of massive galaxies.
While a small number of halos could host a large number of massive galaxies if the majority of the galaxies are satellites, such a HOD results in 
very strong ACF amplitudes at small scales, which conflicts with the present measurements (hence is not selected in the above fittings).
The case (i) models provide a comparable number of galaxies to the observed by including the host halos whose masses are presumably lower 
than in reality, hence the predicted ACFs are weaker than the observed.
Therefore the derived halo masses and biases in the case (i) should be regarded as the lower limits.
The case (ii) takes the opposite approach by selecting only the most massive halos in order to reproduce the observed strong ACFs, while setting aside
the majority of the observed galaxies presumably hosted by lower-mass halos.
Thus the derived halo masses and biases should be the strict upper limits.
We will discuss this issue further below.
The actual HOD parameter values of the galaxies should be somewhere in between those of the above two extreme cases, and we conservatively adopt the above 
upper and lower limits in this work.

\begin{figure*}
  \includegraphics[width=144mm]{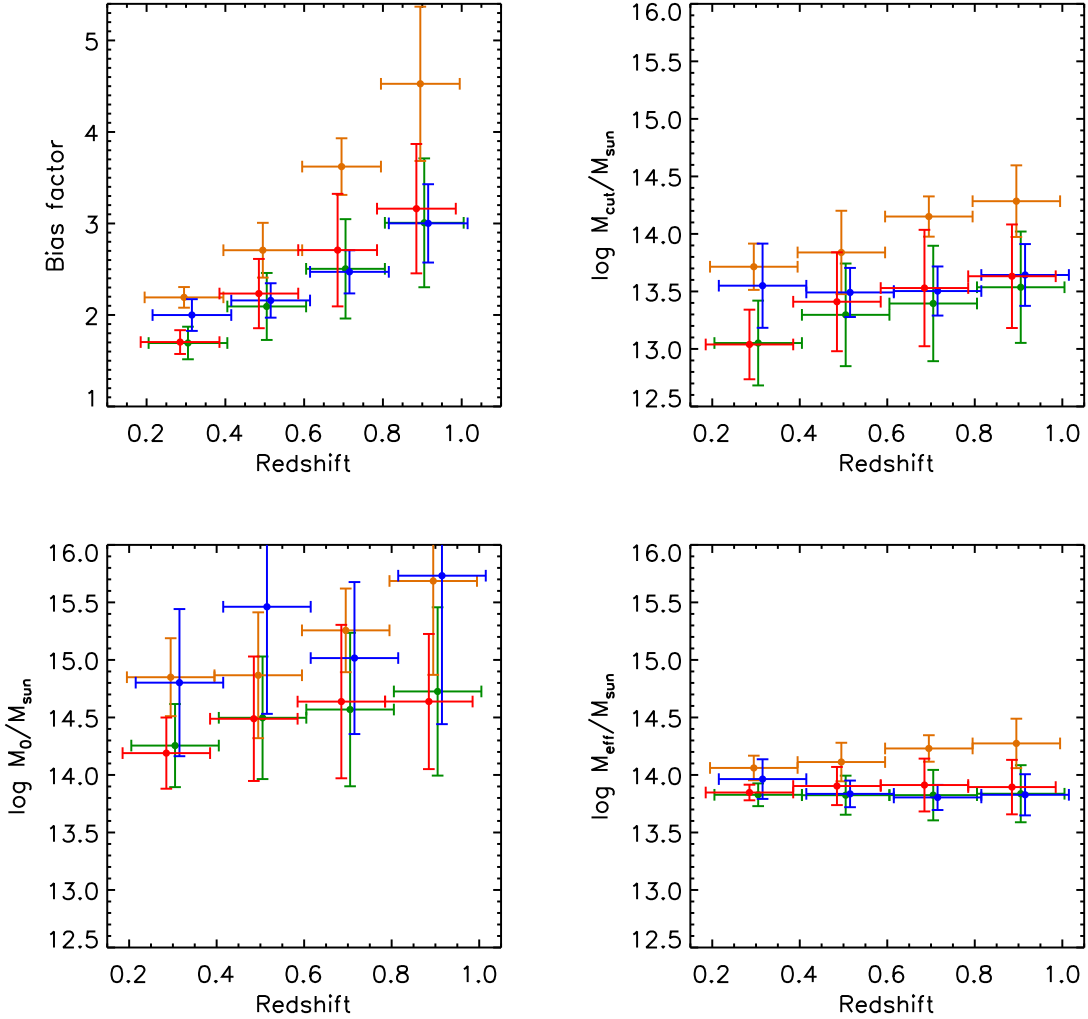}
  \caption{Halo occupation distribution of massive galaxies as a function of redshift 
    (top left: $b_{\rm gal}$, the bias factor, 
    top right: $M_{\rm cut}$, the threshold mass for hosting a central galaxy, 
    bottom left: $M_0$, the typical mass for hosting a satellite galaxy,
    bottom right: $M_{\rm eff}$, the effective mass of host halos).
    The green, orange, blue, and red bars represent the $10^{11.0-11.5} M_{\odot}$, the $10^{11.5-12.0} M_{\odot}$, 
    the blue, and the red galaxies, respectively.}
  \label{hod}
\end{figure*}

In Fig. \ref{hod} we show the derived bias factors and halo mass estimates as a function of redshift.
It is clearly seen that the most massive galaxies mark the most biased density structure of dark matter distributions, as expected.
The threshold mass for hosting a central galaxy, $M_{\rm cut}$, is significantly larger for more massive galaxies, which is consistent with
the general agreement that the capability of hosting massive galaxies depends strongly on halo mass.
The typical mass for hosting a satellite galaxy, $M_0$, is extremely large ($\sim 10^{14-16} M_{\odot}$), which means that only the most
massive halos could host a galaxy with the stellar mass exceeding $10^{11} M_{\odot}$ as a satellite.
Actually we find that the fractions of central galaxies $f_{\rm cen}$ in the current sample are larger than 90 \% in most cases.
These results are the consequence of the relative weakness of the observed ACFs at small scales (one-halo terms, arising from central - satellite pairs) 
compared to those at large scales (two-halo terms, mainly arising from central - central pairs in the current case).
While the effective halo mass $M_{\rm eff}$ shows a similar trend to $M_{\rm cut}$ with regard to the galaxy stellar mass, the difference of 
the $M_{\rm eff}$ between the $10^{11.0-11.5} M_{\odot}$ and the $10^{11.5-12.0} M_{\odot}$ galaxies are less than those of the $M_{\rm cut}$.
It is because of the negative and steepening slope of the halo mass function toward high-mass end; increasing $M_{\rm cut}$ is accompanied by the 
smaller increment of $M_{\rm eff}$ since there are less and less halos as the halo mass increases.
The stellar-mass to halo-mass ratios are found to be $M_{\star}$/$M_{\rm eff} \sim 0.003$.
We observe no significant difference of the ratios between the different stellar-mass classes, while the ratios might be very marginally larger for 
more massive systems, which indicates that the fractional mass growth rate ($\dot{M}/M$) of halos are comparable to or marginally lower than those of 
stellar systems above $M_{\star} \sim 10^{11} M_{\odot}$ (i.e., $\dot{M}/M \la $ $\dot{M_{\star}}/M_{\star}$).
The major mass accretion would have already terminated for these very massive halos, while the stellar-mass assembly of the massive galaxies
would also be inefficient since they are located near the high mass end of the galaxy mass function.

We note that the derived effective masses of halos, $M_{\rm eff} \sim 10^{14} M_{\odot}$, are comparable to the halo masses found 
for galaxy clusters \citep[e.g.,][]{lin04,lin04b}.
Considering the high fraction of central galaxies ($f_{\rm cen} \ga 0.9$), the massive galaxies studied here are presumably equivalent to 
central galaxies of galaxy clusters.
Actually, the stellar masses of $M_{\star} > 10^{11} M_{\odot}$ are typical of the central galaxies of galaxy groups or clusters
with the halo masses $\sim 10^{14} M_{\odot}$ \citep{yang08}, which is comparable to the current estimates of the effective halo mass.

Fig. \ref{hod} shows that the bias factor clearly decreases with decreasing redshift.
It implies rapid evolution of dark matter distributions while the clustering structures traced by the massive galaxies evolve relatively slowly.
The halo mass $M_{\rm cut}$ also shows a marginal decline toward the local universe, which might suggest that the $M_{\star} > 10^{11} M_{\odot}$ galaxies 
are formed in progressively less massive halos with cosmic time.
Then it might imply that the fractional mass growth rate of stellar systems exceeds those of halos ($\dot{M}/M < $ $\dot{M_{\star}}/M_{\star}$)
for the relatively low-mass ($M_{\star} < 10^{11} M_{\odot}$) galaxies, a part of which eventually evolve into the massive galaxies.

We observe no clear difference of the HODs between the blue and the red galaxies in Fig. \ref{hod}.
It contrasts with the measured ACF amplitudes plotted in Fig. \ref{Aw}, where the red galaxies clearly show higher degree of clustering than the blue galaxies.
The difference between the two populations disappears in the derived HODs due to the relative scarcity of the massive blue galaxies;
since the current HOD models predict that both populations are mostly central galaxies and assume that the probability of hosting a central galaxy is a 
monotonically-increasing
function of halo mass, the numerous red central galaxies require relatively low-mass halos as their hosts compared to the scarce blue galaxies.
Thus the derived halo masses (and biases) of the red galaxies become relatively low, which cancels out the higher ACF amplitudes measured for the 
red galaxies (indeed the larger halo masses or biases are predicted for the redder galaxies in the case (ii) which imposes no constraints on the predicted galaxy
number density).

However, the HOD assumption that the probability of hosting galaxies is a monotonically-increasing function of halo mass could be too simplistic particularly
for massive blue galaxies.
It is possible that galaxies in massive, grown-up halos have exhausted most of their gas and have systematically lower level of star formation.
In addition, recent studies suggest star-formation quenching mechanisms of massive galaxies as a feedback process of galaxy mass assembly, such as the 
onset of active galactic nuclei \citep[e.g.,][]{silk98,granato04,springel05}.
These processes could systematically reduce the number of massive blue galaxies in massive halos, in a way dependent on halo characteristics and
galaxy star-formation and its quenching physics.
We also note that fitting the ACFs of the blue and red galaxies separately with the current HOD models might be problematic, since the halos hosting a red 
central and blue satellites, or a blue central and red satellites, cannot be properly reproduced.
These situations might be mimicked by the HOD forms such as those given by the equation (\ref{eq:hodsix}), which allows an upper cut-off in $N_{\rm cen}$
and reproduces the HODs of 'young' galaxies presented by \citet{zheng05},
but we do not obtain improved fits of the ACFs with this form as compared to those with our standard model.
However, since the HODs of the massive blue galaxies could be significantly altered by considering the more numerous red populations, 
the central fractions of massive blue galaxies deduced above are rather tentative.
Further investigation of this issue is beyond the scope of this paper, and would require much more precise clustering measurements of massive blue galaxies.
If the observed scarcity of the massive blue galaxies is due to the star-formation quenching physics rather than their need for very massive host halos, 
then one might be able to conclude, simply from Fig. \ref{Aw}, that massive blue galaxies reside in less massive halos than the hosts of massive red galaxies.
Again, it would be the natural result if redder galaxies have started stellar-mass assembly earlier within the region where the halo mass
assembly has also started earlier.

Finally, we comment on the discrepancy between the HOD model predictions and the observed quantities.
As stated above, the observed ACFs are so strong that very massive halos are required to reproduce them, while such massive halos are extremely 
rare and could not host the observed number of massive galaxies.
It is unlikely that most of the massive galaxies are hosted by a small number of such massive halos as satellites, since it would result in much 
stronger ACFs than observed at small angular scales.
The discrepancy between the observed and model ACFs in the case (i) 
tends to increase with increasing redshift (see Figs. \ref{acf_results} and \ref{acf_results_color}).
In this regard, we note that a similar discrepancy is observed in the analysis of distant red galaxies at $z \sim 2.3$ by \citet{tinker10} (see 
also \citet{quadri08}).
The systematic difference between their model and observed ACFs at large angular scales is roughly a factor of two (while within the 1 $\sigma$ uncertainty 
of their measurements), which is comparable to those found in our sample.
As raised by \citet{quadri08}, the underlying cause of the problem might be that the current HOD models are too simplistic.
While the current models assume that a number of galaxies hosted in a halo depends solely on the halo mass, other halo characteristics such as 
the mass accretion rate could affect the galaxy formation and alter the observed properties of galaxies within the halo.
We also note that fitting the clusterings of different galaxy populations (e.g., blue and red galaxies) at the same time may improve the situation,
while such an analysis would significantly increase the number of HOD parameters and require much precise data for clustering measurements.
Furthermore, we can raise the possibility that the evolution of halo mass function and/or bias function is not well understood; our results might point to
the earlier emergence of massive halos (i.e., larger numbers of massive halos at high redshifts) than predicted in the current halo models.

It is also worth pointing out that the observed ACFs are apparently larger than the fitted HOD models at the smallest angular scales 
($\theta \sim 0.001$ deg).
These close pairs may represent interacting galaxies \citep[e.g.,][]{roche99}.
Reproducing such a feature by the HOD models requires the detailed description of galaxy interaction within a halo.
However, our focus here is the global characteristic of the host halos and such an investigation is a subject of future papers.

\section{Summary}

In this work we present a clustering analysis of $\sim$60,000 massive (stellar mass $M_{\star} > 10^{11} M_{\odot}$) galaxies taken from \citet{matsuoka10}.
The galaxies have been extracted from 55.2 deg$^2$ of the UKIRT Infrared Deep Sky Survey (UKIDSS) and the Sloan Digital Sky Survey (SDSS) II 
Supernova Survey, and are nearly complete to $z = 1$ with robust estimates of photometric redshift and stellar mass.
We classify them based on stellar masses ($M_{\star} = 10^{11.0-11.5} M_{\odot}$ or $10^{11.5-12.0} M_{\odot}$) and rest-frame colors (blue: $U - V < 1.0$, 
red: $U - V > 1.0$), in order to reveal the difference in the spatial distributions of different galaxy populations.

The angular correlation functions (ACFs) of the galaxies are quantified using the \citet{landy93} estimator, and
we find strong clustering detected for all the subsamples.
In order to interpret the measured ACFs, we construct halo occupation distribution (HOD) models following \citet{blake08} and compare the predicted ACFs
with the observed.
Our major findings are as follows.
\begin{enumerate}
\item The clustering amplitudes clearly show a systematic trend regardless of the measured redshift, in which the most massive, $10^{11.5-12.0} M_{\odot}$ 
galaxies have the strongest clustering and the blue galaxies have the weakest clustering.
It would be the natural result if more mature galaxies have started stellar-mass assembly earlier within the highly-biased region where the halo mass 
assembly has also started earlier.
\item The bias factors and halo masses are systematically larger for the galaxies with larger stellar masses, which confirms that the capability of hosting
massive galaxies depends strongly on halo mass.
The stellar-mass to halo-mass ratios are $M_{\star}$/$M_{\rm eff} \sim 0.003$, with no significant difference observed between the different stellar-mass classes.
\item The inferred halo masses of $M_{\rm eff} \sim 10^{14} M_{\odot}$ and the high central fractions ($f_{\rm cen} \ga$ 0.9) indicate that the massive
galaxies in the current sample are equivalent to central galaxies of galaxy clusters.
\item The bias factor decreases with decreasing redshift, which implies rapid evolution of dark matter distributions while the clustering structures
traced by the massive galaxies evolve relatively slowly.
The halo mass $M_{\rm cut}$ also shows a marginal decline toward the local universe, which might suggest that massive ($M_{\star} > 10^{11} M_{\odot}$)
galaxies are formed in progressively less massive halos with the cosmic time.
\item If the observed scarcity of massive blue galaxies is due to the star-formation quenching physics rather than their need for very massive host halos, 
then one might be able to conclude that massive blue galaxies reside in less massive halos than the hosts of massive red galaxies.
\item The observed ACFs are so strong at large angular scales that very massive halos are required to reproduce them, while such massive halos are extremely
rare and could not host the observed number of massive galaxies.
It might point to the inadequacy of the current HOD models or a lack of knowledge about the evolving halo mass function and/or bias function.
\end{enumerate}

\section*{Acknowledgments}

We are grateful to K. Shimasaku, K. Kohno, J. Makino, N. Yasuda, and N. Yoshida for insightful discussions and suggestions.
We thank the referee for his/her useful comments which helped to improve this paper.
YM acknowledges Grant-in-Aid from the Research Fellowships of the Japan Society for the Promotion of Science (JSPS) for Young Scientists. 
This work was supported by Grants-in-Aid for Scientific Research (17104002, 21840027, 22684005),  Specially Promoted Research (20001003),
Specially Promoted Research on Inovated Area (22111503), and the Global COE Program of Nagoya University "Quest for Fundamental Principles 
in the Universe (QFPU)" from JSPS and MEXT of Japan.

\label{lastpage}

\end{document}